\documentclass[a4paper,11pt]{article}
\usepackage{pos}
\usepackage{amsmath,mathtools,bbold}

\usepackage{float,wrapfig,subcaption}

\title{Asymptotic lattice spacing dependence of spectral quantities in lattice QCD with Wilson or Ginsparg-Wilson quarks}
\ShortTitle{Lattice spacing dependence of spectral quantities in LQCD with Wilson or GW quarks}

\author*[a]{Nikolai Husung}

\affiliation[a]{Physics and Astronomy, University of Southampton,\\ Southampton SO17 1BJ, UK}

\emailAdd{n.husung@soton.ac.uk}

\abstract{One major systematic uncertainty of lattice QCD results is due to the continuum extrapolation.
For an asymptotically free theory like QCD one finds corrections of the form $a^{n_\mathrm{min}}[2b_0\bar{g}^2(1/a)]^{\hat{\Gamma}_i}$ with lattice spacing $a$, where $\bar{g}(1/a)$ is the running coupling at renormalisation scale $\mu=1/a$ and $n_\mathrm{min}$ is a positive integer.
$\hat{\Gamma}_i$ can take any positive or negative value, but is computable by next-to-leading order perturbation theory.
It will impact convergence towards the continuum limit.
Balog, Niedermayer and Weisz first pointed out how problematic such corrections can be in their seminal work for the O(3) model.

Based on Symanzik Effective Theory for lattice QCD with Ginsparg-Wilson and Wilson quarks, various powers $\hat{\Gamma}_i$ are found due to lattice artifacts from the discretised lattice action.
Those powers are sufficient when describing spectral quantities, while non-spectral quantities will require additional powers originating from corrections to each of the discretised local fields involved.
This new input should be incorporated into ans\"atze used for the continuum extrapolation.}

\FullConference{%
The 39th International Symposium on Lattice Field Theory,\\
8th-13th August, 2022,\\
Rheinische Friedrich-Wilhelms-Universität Bonn, Bonn, Germany
}

\def\nmin{n_\mathrm{min}}
\def\gbar{\bar{g}}
\def\ord{\mathrm{O}}
\def\MSbar{\ensuremath{\overline{\text{MS}}}}
\def\op{\mathcal{O}}
\def\opE{\mathcal{E}}
\def\base{\mathcal{B}}
\def\L{\mathscr{L}}
\def\Nf{N_\mathrm{f}}
\def\obs{\mathcal{P}}
\def\rmd{\mathrm{d}}
\def\tr{\mathop{\mathrm{tr}}}
\def\diag{\mathop{\mathrm{diag}}}

\def\opFive{\op^{(1)}}
\def\opSix{\op^{(2)}}
\def\opmin{\op^{(\nmin)}}

\def\pagelegend{10}
\def\cGammaFile{cGammaHnoNf.pdf}

\def\cred{\color{red}}
\newcommand{\cev}[1]{\overset{\leftarrow}{#1}}

\begin{document}
\maketitle

\restylefloat{figure}

\section{Introduction}
When aiming at precision physics predictions from lattice QCD, all systematic uncertainties must be under control.
Although lattice spacings down to $a\simeq 0.04\,\mathrm{fm}$ are accessible nowadays for the lattice discretisations considered here, see e.g.~\cite{Boyle:2017jwu,Bruno:2017gxd}, the continuum extrapolation $a\searrow 0$ still remains an important source of systematic uncertainties.
The aim of the investigation~\cite{Husung:2019ytz,Husung:2021tml,Husung:2021mfl} presented here is to work out the leading asymptotic lattice-spacing dependence expected for spectral quantities, when using one of the following lattice-fermion discretisations:
\begin{itemize}
\item Wilson quarks~\cite{Wilson:1974,Wilson:1975id} with and without non-perturbative $\ord(a)$ improvement~\cite{Luscher:1996ug},
\item Ginsparg-Wilson~(GW) quarks~\cite{Ginsparg:1981bj} using Overlap fermions~\cite{Neuberger:1997fp,Neuberger:1998wv} as an example.
This also includes Domain-Wall fermions (DWF)~\cite{Kaplan:1992bt,Furman_1995,Brower:2012vk} when assuming an (approximate) lattice chiral symmetry~\cite{Luscher:1998pqa} in the massless case.
\end{itemize}

\begin{wrapfigure}{r}{0.46\textwidth}
\includegraphics[scale=1,page=1]{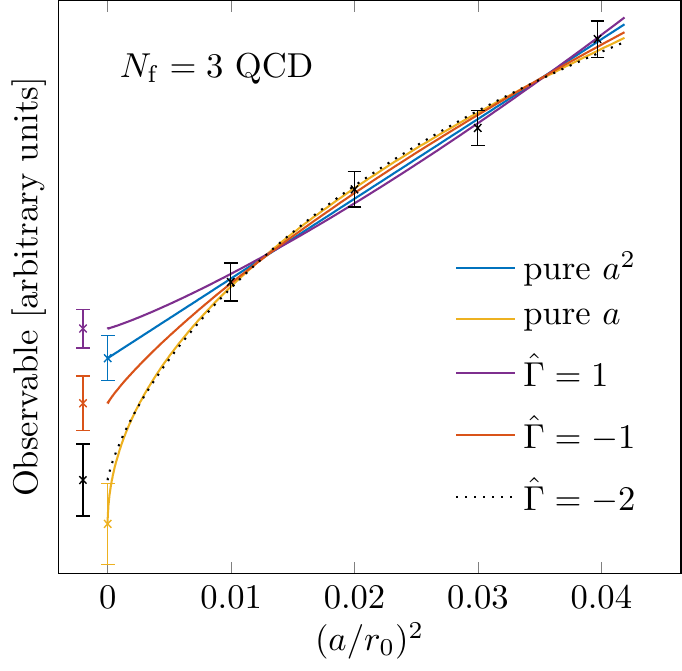}
\caption{Continuum extrapolation of fake data points randomly generated from eq.~\eqref{eq:fakeData}.
The various ans\"atze extrapolating to a constant include the typical choice of simple $a^2$ corrections and unconventional powers in the coupling $\hat{\Gamma}\in\{-2,-1,1\}$ modifying $a^2[2b_0\gbar^2(1/a)]^{\hat{\Gamma}}$, as well as simple $a$ corrections.
Both $a$ and (the correct ansatz) $a^2[2b_0\gbar^2(1/a)]^{-2}$ agree within uncertainties, while the simple $a^2$ by construction completely misses the curvature.
Three continuum extrapolated values have been shifted to the left for better readability.\\[-3pt]}
\label{fig:exampleFakeData}
\end{wrapfigure}
Based on symmetries realised by the lattice theory and analysis of canonical mass-dimensions the leading lattice artifacts can be written as positive integer powers $a^{\nmin}$ in the lattice spacing.
While this picture is sufficient for a classical field theory, quantum effects will modify the leading lattice-spacing dependence.
In an asymptotically free theory like QCD, the leading lattice artifacts instead take the form $a^{\nmin}[2b_0\gbar^2(1/a)]^{\hat{\Gamma}_i}$, where $\hat{\Gamma}_i$ are related to the 1-loop anomalous dimensions of $(4+\nmin)$-dimensional ``irrelevant'' operators in a Symanzik Effective Field Theory~(SymEFT)~\cite{Symanzik:1979ph,Symanzik:1981hc,Symanzik:1983dc,Symanzik:1983gh}.
Those $\hat{\Gamma}_i$ can take \emph{arbitrary} real values.
How important this effect can be, was first shown in the seminal work of Balog, Niedermayer and Weisz~\cite{Balog:2009np,Balog:2009yj} for the O(3) model.
There one finds $\min_i\hat{\Gamma}_i=-3$, which has a huge impact on the convergence towards the continuum limit.
To highlight what impact this could have in 3-flavour QCD, figure~\ref{fig:exampleFakeData} shows Gaussian fake data points randomly generated as the function
\begin{equation}
f(a)=a^2[2b_0\gbar^2(1/a)]^{-2}\label{eq:fakeData},
\end{equation}
i.e.~$\nmin=2$ and a more conservatively chosen $\hat{\Gamma}=-2$.
We use the 5-loop running coupling in the $\overline{\text{MS}}$ scheme~\cite{Luthe:2017ttc}.
While having only one power in the coupling is an oversimplification, already this example shows that a simple $a^2$ ansatz may not yield the desired continuum-extrapolated value within uncertainties.
From another point of view, one is underestimating the systematic uncertainty of the continuum extrapolation.
In this example, the actual $a$-dependence looks even more similar to a simple $a^1$ ansatz.
For real data points one will have not only more powers in the coupling at $\ord(a^{\nmin})$ but higher order corrections $\ord(a^{\nmin+1})$ may also play a role.
Having an idea what the leading powers $\hat{\Gamma}_i$ are is therefore advisable to guide ans\"atze used for continuum extrapolations and incorporate them into the estimate of the associated systematic uncertainty.

\section{Symanzik Effective Field Theory}
The leading lattice artifacts can be described in terms of a SymEFT given by the Lagrangian
\begin{equation}
\L(x)=\L_{\mathrm{QCD}}(x)+a^{\nmin}\sum_i\omega_i(g_0^2)\opmin_i(x)+\ord(a^{\nmin+1}),
\end{equation}
where $\omega_i$ are the bare matching coefficients, and
\begin{equation}
\L_{\mathrm{QCD}}=-\frac{1}{2g_0^2}\tr(F_{\mu\nu}F_{\mu\nu})+\bar{\Psi}\left(\gamma_\mu D_\mu(A)+M\right)\Psi
\end{equation}
is the Lagrangian of continuum QCD with $\Nf$ flavours, $D_\mu(A)=\partial_\mu+A_\mu$, $A_\mu\in\mathrm{su}(N)$ and $F_{\mu\nu}=[D_\mu,D_\nu]$.
The local operators $\op_i^{(\nmin)}(x)$ have canonical mass-dimension $(4+\nmin)$  and form a minimal basis describing all lattice artifacts arising from the discretised lattice action to $\ord(a^{\nmin})$.
Which operators can occur is constrained by the symmetries realised by the lattice formulation, here
\begin{itemize}
\item Local SU($N$) gauge symmetry,
\item invariance under charge conjugation and any Euclidean reflection,
\item Hypercubic symmetry as a remnant of broken O(4) symmetry,
\item $\mathrm{SU}(\Nf)_\mathrm{L}\times\mathrm{SU}(\Nf)_\mathrm{R}\times\mathrm{U}(1)_\mathrm{V}$ flavour symmetry for massless lattice-fermion actions preserving lattice chiral symmetry,
\item $\mathrm{U}(\Nf)_\mathrm{V}$ flavour symmetry for massless (or mass-degenerate) Wilson quarks.
\end{itemize}
For on-shell physics in the sense of~\cite{Luscher:1996sc} we can make use of the continuum equations of motion~(EOMs)
\begin{equation}
[D_\mu(A),F_{\mu\nu}]=g_0^2T^a\bar{\Psi}\gamma_\nu T^a\Psi,\quad \gamma_\mu D_\mu\Psi=-M\Psi,\quad \bar{\Psi}\cev{D}_\mu\gamma_\mu=\bar{\Psi}M,
\end{equation}
to reduce the minimal basis further to form an \emph{on-shell} minimal basis.
Here $T^a$ denotes a generator of the su$(N)$ algebra with normalisation $\tr(T^aT^b)=-\delta^{ab}/2$.

The minimal on-shell basis at mass-dimension~5 relevant for unimproved Wilson quarks reads~\cite{Sheikholeslami:1985ij}
\begin{align}
\opFive_1&=\frac{i}{4}\bar{\Psi}\sigma_{\mu\nu}F_{\mu\nu}\Psi,& \opFive_2&=\frac{\tr(M)}{g_0^2}\tr(F_{\mu\nu}F_{\mu\nu}),&\opFive_3&=\bar{\Psi}M^2\Psi,\vphantom{\frac{\tr(M^2)}{g_0^2}}\nonumber\\
\opFive_4&=\tr(M)\bar{\Psi}M\Psi,&\opFive_5&=\tr(M^2)\bar{\Psi}\Psi,&\opFive_6&=\tr(M)^2\bar{\Psi}\Psi,\vphantom{\frac{\tr(M^2)}{g_0^2}}\label{eq:dim5ops}
\end{align}
where $\sigma_{\mu\nu}=\frac{i}{2}[\gamma_\mu,\gamma_\nu]$.
The full mass-dimension~6 on-shell operator basis considered here is
\begin{align}
\opSix_1&=\frac{1}{g_0^2}\tr(D_\mu F_{\nu\rho}D_\mu F_{\nu\rho}),&\opSix_2&=\frac{1}{g_0^2}\sum_\mu\tr(D_\mu F_{\mu\nu}D_\mu F_{\mu\nu}),\vphantom{\frac{\tr(M^2)}{g_0^2}}\nonumber\\
\opSix_3&=\sum_\mu\bar{\Psi}\gamma_\mu D_\mu^3\Psi,&\opSix_{4\leq i\leq 13}&=g_0^2(\bar{\Psi}\Gamma_i\Psi)^2,\vphantom{\frac{\tr(M^2)}{g_0^2}}\nonumber\\
\opSix_{14}&=\frac{i}{4}\bar{\Psi}M\sigma_{\mu\nu}F_{\mu\nu}\Psi,&\opSix_{15}&=\frac{\tr(M^2)}{g_0^2}\tr(F_{\mu\nu}F_{\mu\nu}),\vphantom{\frac{\tr(M^2)}{g_0^2}}\nonumber\\
\opSix_{16}&=\bar{\Psi}M^3\Psi,&\opSix_{17}&=\tr(M^2)\bar{\Psi}M\Psi,\vphantom{\frac{\tr(M^2)}{g_0^2}}\nonumber\\
\opSix_{18}&=\frac{i\tr(M)}{4}\bar{\Psi}\sigma_{\mu\nu}F_{\mu\nu}\Psi,& \opSix_{19}&=\frac{\tr(M)^2}{g_0^2}\tr(F_{\mu\nu}F_{\mu\nu}),\vphantom{\frac{\tr(M^2)}{g_0^2}}\nonumber\\
\opSix_{20}&=\tr(M)\bar{\Psi}M^2\Psi,&
\opSix_{21}&=\tr(M)^2\bar{\Psi}M\Psi,\vphantom{\frac{\tr(M^2)}{g_0^2}}\nonumber\\
\opSix_{22}&=\tr(M^3)\bar{\Psi}\Psi,&
\opSix_{23}&=\tr(M)\tr(M^2)\bar{\Psi}\Psi,\vphantom{\frac{\tr(M^2)}{g_0^2}}\nonumber\\
\opSix_{24}&=\tr(M)^3\bar{\Psi}\Psi,\vphantom{\frac{\tr(M^2)}{g_0^2}}
\end{align}
where $\Gamma_{4\leq i\leq 7}\in\{\gamma_\mu,\gamma_\mu\gamma_5\}\otimes\{\mathbb{1},T^a\}$ and $\Gamma_{8\leq i\leq 13}\in\{\mathbb{1},\gamma_5,i\sigma_{\mu\nu}\}\otimes\{\mathbb{1},T^a\}$.
The basis listed here is on-shell equivalent to the one found in~\cite{Sheikholeslami:1985ij}.
The number of 4-quark operators needed depends on whether lattice chiral symmetry is realised in the massless case or not, i.e., the operators $\opSix_{8\leq i\leq 13}$ only contribute in the case of Wilson quarks.
To see this it suffices to rewrite the mass term
\begin{equation}
\bar{\Psi}M\Psi = \bar{\Psi}_\mathrm{R}M\Psi_\mathrm{L}+\bar{\Psi}_\mathrm{L}M^\dagger\Psi_\mathrm{R},
\end{equation}
where the subscripts indicate left- and right-handed quarks respectively.
To apply this reasoning to the GW lattice action the projection to left- and right-handed quarks must be done with the modified $\hat\gamma_5=\gamma_5\left\{1+\frac{a}{2}\hat{D}_\mathrm{GW}\right\}$.
Promoting the quark-mass matrix to a spurionic field, we find that flavour rotations $L\in \mathrm{SU}(\Nf)_\mathrm{L}$ and $R\in \mathrm{SU}(\Nf)_\mathrm{R}$
\begin{equation}
\bar{\Psi}_\mathrm{L}\rightarrow \bar{\Psi}_\mathrm{L}L^\dagger,\quad \bar{\Psi}_\mathrm{R}\rightarrow \bar{\Psi}_\mathrm{R}R^\dagger,\quad \Psi_\mathrm{L}\rightarrow L\Psi_\mathrm{L},\quad \Psi_\mathrm{R}\rightarrow R\Psi_\mathrm{R},
\end{equation}
keep the GW lattice action and therefore the corresponding SymEFT on-shell action invariant if the (spurionic) mass matrix transforms like
\begin{equation}
M\rightarrow RML^\dagger.
\end{equation}
This allows one to exclude the massive operators $\opSix_{i\geq18}$ for GW quarks as they are incompatible to this spurionic symmetry.\footnote{I thank the referee of~\cite{Husung:2022kvi} for pointing out that inherent structure.
While it does not make a difference regarding the spectrum of powers in the coupling $\hat{\Gamma}_i$ found here, it clearly shrinks the number of operators allowed in the case of GW quarks.}
Prepending an overall power $g_0^2$ to the 4-quark operators is the natural choice given the leading order at which 4-quark interactions can occur earliest in lattice perturbation theory, corresponding to connected tree graphs.
Instead of the connected graphs we may just as well consider the one-particle-irreducible pieces.
We thus immediately find that all tree-level matching coefficients of the 4-quark operators in this basis vanish except for contributions arising from the use of the gluonic EOM.

\section{Leading powers \boldmath$\hat\Gamma_i$ in the coupling}
If we consider a Renormalisation Group invariant (RGI) spectral quantity $\obs(a)$ such as a hadron mass computed on a lattice with lattice spacing $a$, its leading lattice artifacts are
\begin{equation}
\frac{ \obs(a)}{\lim\limits_{a\searrow 0} \obs(a)}=1-a^{\nmin}\sum_ic_i^\op(\gbar^2(1/a),a\mu)\delta\obs_i^\op(\mu)+\ord(a^{\nmin+1}),\label{eq:dlatt}
\end{equation}
where $\delta\obs_i^\op$ is the renormalised contribution of the $i$-th operator to the overall lattice artifacts normalised by $\obs(0)$ and $c_i^\op$ denotes the corresponding renormalised matching coefficient at renormalisation scale $\mu=1/a$.
For compactness we will drop the argument $a\mu$, setting $a\mu=1$, of the renormalised matching coefficients.
The remaining scale dependence of $\delta\obs_i(\mu)$ is governed by the Renormalisation Group equation~(RGE)
\begin{equation}
\mu\frac{\rmd\delta\obs_i^\op(\mu)}{\rmd\mu}=-\gbar^2(\mu)\left\{\gamma_0^\op+\ord(\gbar^2(\mu))\right\}_{ij}\delta\obs_j^\op(\mu),
\end{equation}
where $\gamma_0^\op$ are the 1-loop anomalous dimensions of our minimal operator basis.
To obtain $\gamma_0^\op$ we then need to compute how the operators mix under renormalisation to 1-loop order.
\begin{figure}
\def\scale{1}
\begin{subfigure}[t]{0.19\textwidth}\centering
\includegraphics[scale=\scale]{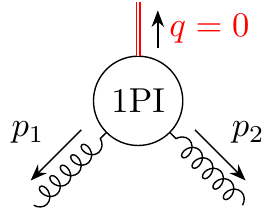}
\end{subfigure}
\begin{subfigure}[t]{0.19\textwidth}\centering
\includegraphics[scale=\scale]{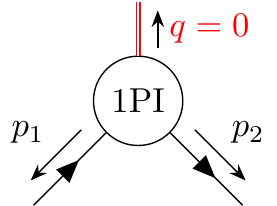}
\end{subfigure}
\begin{subfigure}[t]{0.19\textwidth}\centering
\includegraphics[scale=\scale]{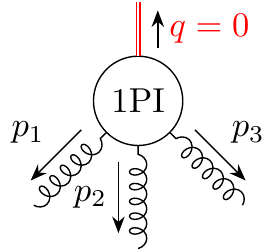}
\end{subfigure}
\begin{subfigure}[t]{0.19\textwidth}\centering
\includegraphics[scale=\scale]{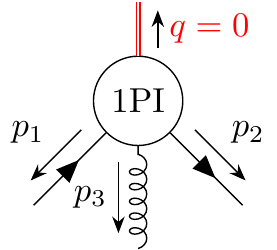}
\end{subfigure}
\begin{subfigure}[t]{0.19\textwidth}\centering
\includegraphics[scale=\scale]{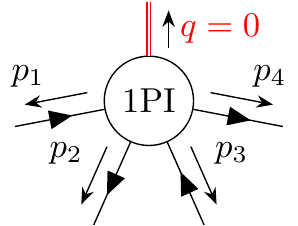}
\end{subfigure}
\caption{1PI graphs with operator insertion $\tilde{\op}_i^{(d)}({\cred q=0})$ considered to compute the 1-loop operator mixing.}\label{fig:1PIgraphs}
\end{figure}
The 1-loop renormalisation has been determined by computing the 1PI graphs depicted in figure~\ref{fig:1PIgraphs} in background field gauge~\cite{tHooft:1975uxh,Abbott:1980hw,Abbott:1981ke,Luscher:1995vs} and working in dimensional regularisation~\cite{tHooft:1972tcz,tHooft:1973mfk} in $D=4-2\epsilon$ dimensions.
Since we are interested only in the 1-loop anomalous dimensions and we work in the mass-independent $\MSbar{}$ renormalisation scheme~\cite{Bardeen:1978yd}, we can make use of so-called infrared-rearrangement, see e.g.~\cite{Misiak:1994zw,Chetyrkin:1997fm,Luthe:2017ttc}, by rewriting
\begin{equation}
\frac{1}{(k-p)^2+M^2}=\frac{1}{k^2+\Omega}-\frac{2kp+p^2+M^2-\Omega}{[k^2+\Omega][(k-p)^2+M^2]},
\end{equation}
where $k$ is the loop momentum and $\Omega>0$ is an arbitrary constant.
The second term on the right hand side is less UV divergent.
Iterating this step, the UV divergent part can be written solely in terms of $\int\rmd^Dk\,k_{\mu_1}\ldots k_{\mu_l}/[k^2+\Omega]^{n}$, where $l,n\in\mathbb{N}\cup\{0\}$.
Apart from the operator insertion, we keep the external momenta of the 1PI graphs free.
This allows mixing of off-shell contributions from operators $\opE^{(d)}$ vanishing by EOMs
\begin{equation}
\begin{pmatrix}
\op_i \\
\opE_j
\end{pmatrix}_{\MSbar{}}^{(d)}=\begin{pmatrix}
Z_{ik}^\op & Z_{il}^{\op,\opE} \\
0 & Z_{jl}^\opE
\end{pmatrix}
\begin{pmatrix}
\op_k \\
\opE_l
\end{pmatrix}^{(d)}.
\end{equation}
Contributions of the operators $\opE^{(d)}$ vanish for on-shell physics, hence the triangular form of the mixing matrix.
We are interested only in the on-shell part $Z^\op$, from which we finally obtain the anomalous-dimension matrix
\begin{equation}
\gamma^\op(\gbar^2(\mu))=\mu\frac{\rmd Z^\op}{\rmd\mu}(Z^\op)^{-1}=-\gbar^2(\mu)\left\{\gamma_0^\op + \ord(\gbar^2(\mu))\right\}.
\end{equation}

Eventually we can bring the 1-loop anomalous-dimension matrix into diagonal form\footnote{If $\gamma_0^\op$ is non-diagonalisable, one may still bring it into Jordan normal form, where the remaining off-diagonal entries give rise to additional factors $\log(\gbar^2)$ already at leading order.
This becomes relevant when considering quenched or mixed actions.
See \cite{Husung:2022kvi} for the necessary generalisations in this case.} through the change of basis $ \base_i^{(\nmin)}=\nu_{ij}\op_j^{(\nmin)}$ and define
\begin{equation}
\hat{\gamma}=\frac{\diag(\gamma_0^\base)}{2b_0},
\end{equation}
where $b_0$ is the 1-loop coefficient of the QCD $\beta$-function.
After the change of basis we can solve the RGE via
\begin{equation}
\delta \obs_i^\base(\mu)=[2b_0\gbar^2(\mu)]^{\hat{\gamma}}_{ij}\delta\obs_{j;\text{RGI}}^\base\times\{1+\ord(\gbar^2(\mu))\}\,,
\end{equation}
where $\delta\obs_{j;\text{RGI}}^\base$ is RGI and all scale dependence is absorbed into the prefactor.

Next, we need to determine the matching coefficients in the diagonal basis
\begin{equation}
c_i^\base(\gbar^2(1/a))=c_j^\op(\gbar^2(1/a))\nu^{-1}_{ji}=\hat{c}_i[2b_0\gbar^2(1/a)]^{n_i}\times\{1+\ord(\gbar^2(1/a))\}.\label{eq:LOmatchingDiag}
\end{equation}
where $\hat{c}_i$ is a non-zero constant and $n_i\in\mathbb{N}\cup\{0\}$.
The leading power in $\gbar^2(1/a)$ of the $i$-th operator may be further suppressed depending on the value $n_i\geq 0$.
For example $n_i\geq1\forall i$ corresponds to full tree-level improvement.
To take this into account, we introduce
\begin{equation}
\hat{\Gamma}_i=\hat{\gamma}_i+n_i
\end{equation}
as the truly leading power of the $i$-th operator.
Since we are mainly interested in the leading asymptotic lattice-spacing dependence, we are content with tree-level matching.\footnote{This is true unless $\min_i\hat{\gamma}_i$ has a vanishing tree-level matching coefficient, while all other contributions' $\hat{\gamma}_j$ lead to a suppression by at least one power in $\gbar^2(1/a)$.
Here this is not the case, otherwise the matching must be performed to subleading powers.}
Following eq.~\eqref{eq:LOmatchingDiag}, we can determine the TL matching coefficients in the basis $\op_i$ before making the change of basis.
This avoids having to deal with lattice perturbation theory, because matching to this order can be achieved by the classical $a$-expansion of the lattice action.
As a consequence, we know each coefficient $c_i^\base$ only to leading order.
If a coefficient vanishes to leading order, this implies $n_i>0$.
Because we are unaware of what happens at subleading orders, we then assume $n_i=1$ for any vanishing leading-order matching coefficients.

\subsection{Leading asymptotic lattice spacing dependence}
Plugging the expressions derived above back into eq.~\eqref{eq:dlatt} yields the desired formula for the leading asymptotic lattice-spacing dependence
\begin{equation}
\frac{ \obs(a)}{\lim\limits_{a\searrow 0} \obs(a)}=1-a^{\nmin}\sum_i\hat{c}_i[2b_0\gbar^2(1/a)]^{\hat{\Gamma}_i}\delta\obs_{i;\text{RGI}}^\base\times\{1+\ord(\gbar^2(1/a))\}+\ord(a^{\nmin+1})\,.\label{eq:dlattRGI}
\end{equation}
Before we can make use of eq.~\eqref{eq:dlattRGI}, some remarks are necessary on the conventions we use
\begin{enumerate}
\item We normalise each row-vector $v_i$ describing the change of basis $\base_i^{(\nmin)}=v_{ij}\op_j^{(\nmin)}$ such that the largest absolute component is set to one.
This may lead to a different normalisation between the massive and the massless case for the same operator.
\item Operators with degenerate $\hat{\Gamma}_\text{deg}$ (and no logs) cannot be distinguished numerically and are combined as
\begin{equation}
\base_\text{deg}^{(d)}=\frac{1}{\hat{c}_\text{max}}\sum_{i:\,\hat{\Gamma}_i=\hat{\Gamma}_\text{deg}}\hat{c}_i\base_i^{(d)},
\end{equation}
where $\hat{c}_\text{max}$ is the coefficient with largest absolute value being summed over.
\item We \emph{assume} that the various $|\delta\obs_{i;\text{RGI}}|$ have the same order of magnitude despite the somewhat arbitrary normalisation of the basis.
With this assumption in mind, we then compare the overall orders of magnitude of $\hat{c}_i$ and $\hat{c}_j$ as well as the associated powers $\hat{\Gamma}_i$ and $\hat{\Gamma}_j$.
Notice that the normalisation of the operator basis will scale $\delta\obs_{i;\text{RGI}}$ and $\hat{c}_i$ inversely to one another.
\end{enumerate}
All statements made in the following necessarily depend on the chosen normalisation of the operator basis $\base^{(\nmin)}$.
Since $\delta\obs_{i;\text{RGI}}$ are unknown, quantity-dependent constants, we can only give a qualitative picture.

\subsection{Some examples}
\def\scale{0.74}
\begin{figure}\centering
\includegraphics[scale=\scale,page=1]{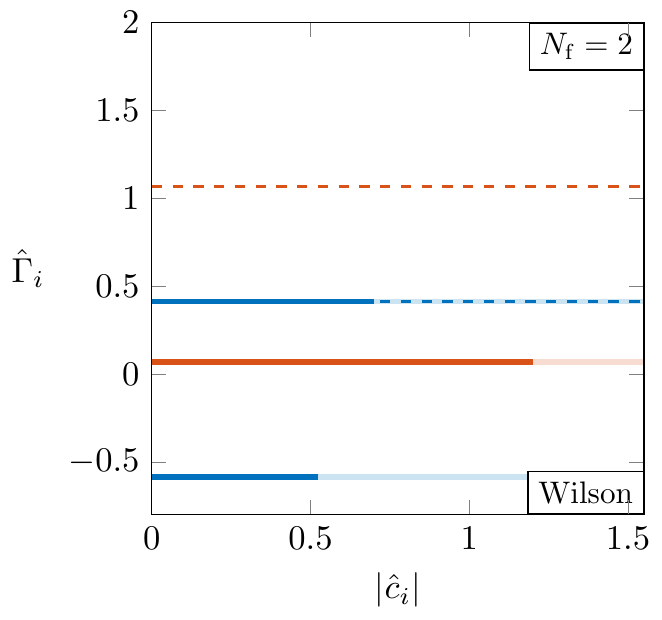}
\includegraphics[scale=0.86,page=\pagelegend,clip=true,trim=0 0 140 0]{addimages/\cGammaFile}
\caption{Leading powers $\hat{\Gamma}_i$ in $\gbar^2(1/a)$ for the $i$-th operator $\base_i^{(1)}$ of Wilson QCD plotted against their associated leading-order matching coefficients $\hat{c}_i$.
At subleading order, the value of the matching coefficient is unknown and instead a dashed line is drawn for the first subleading power, i.e.~at $\hat{\Gamma}_\text{min}+1$.
The faded line is drawn to highlight severely suppressed leading order matching coefficients.}\label{fig:WilsonExample}
\end{figure}
The leading asymptotic lattice-spacing dependence in eq.~\eqref{eq:dlattRGI} is a function of both $\hat{c}_i$ and $\hat{\Gamma}_i$.
To capture this dependence, we consider a two-dimensional plot, where the axis have been flipped in comparison to~\cite{Husung:2021mfl}.
The case of $\ord(a)$ lattice artifacts of unimproved 2-flavour Wilson QCD is depicted in figure~\ref{fig:WilsonExample} and serves as an introduction to this kind of representation, which contains a variety of information
\begin{enumerate}
\item The leading power $\hat{\Gamma}_\text{min}$ indicates whether the overall lattice artifacts are logarithmically enhanced for $\hat{\Gamma}_\text{min}<0$ compared to plain $a^{\nmin}$ or further suppressed $\hat{\Gamma}_\text{min}>0$.
\item Density of the various $\Delta \hat{\Gamma}_{ij}=\hat{\Gamma}_i-\hat{\Gamma}_j$ accounts for the relative suppression in powers of $\gbar^{2\Delta \hat{\Gamma}_{ij}}(1/a)$ for different contributions.
For too small $|\Delta \hat{\Gamma}_{ij}|$ suppression cannot be expected.
\item Absolute values of $|\hat{c}_i|$ are relevant to decide whether some contributions are overall enlarged, possibly undoing any relative suppression in powers $\Delta \hat{\Gamma}_{ij}$.
Severely enlarged coefficients also hint at cases where tree-level improvement may be advisable or where lattice parameters can be tuned in a better way as a function of the bare lattice coupling.
\end{enumerate}
For the case of unimproved 2-flavour Wilson QCD, we find only three distinct powers $\hat{\Gamma}_i$ at leading order, where the lowest power originates from a massive operator with $\hat{\Gamma}_\text{min}\approx-0.59\gg -3$.
Here and in the following the subscript ``min'' used on $\hat{c}_{\text{min}}$ and $\hat{\Gamma}_\text{min}$ will refer to the contribution dominant by power-counting in $\gbar^2(1/a)$.
Since there are only three distinct values, the spectrum of those powers is sufficiently gapped that one might actually be able to decide which contribution dominates in the massive and especially in the (close to) massless case.
All leading-order matching coefficients are roughly the same order of magnitude, meaning that no particular contribution is expected to be strongly enhanced over the others.

\begin{figure}
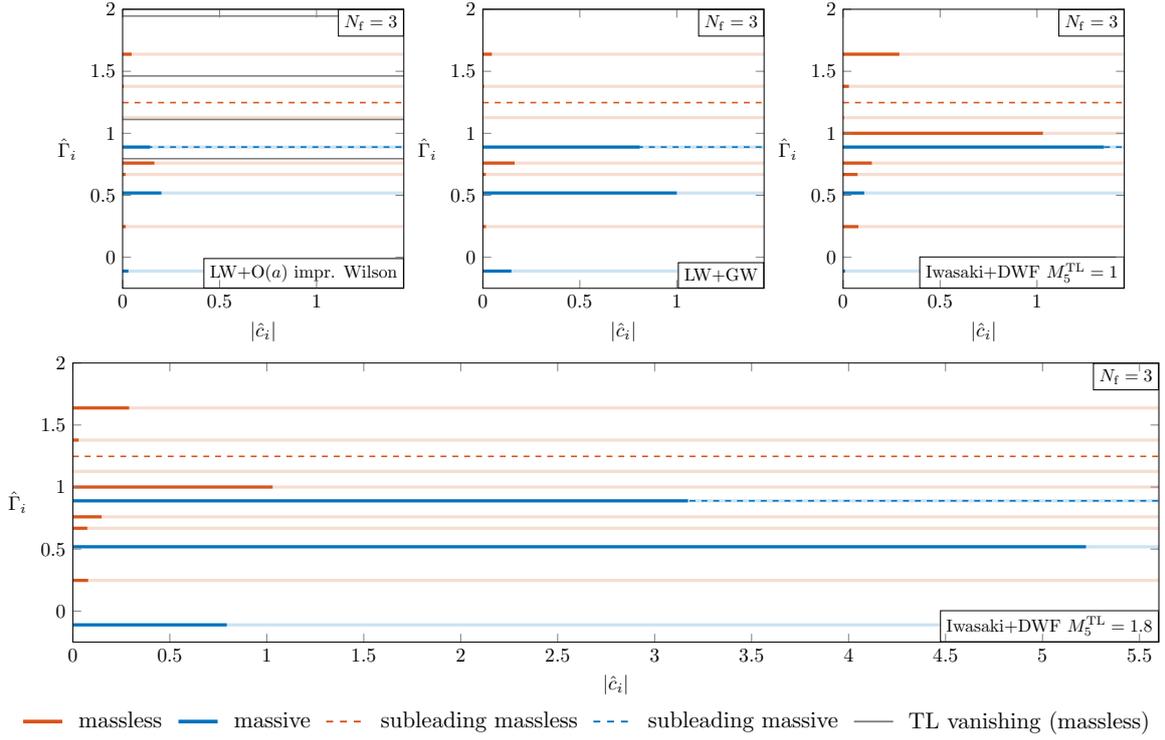
\centering
\includegraphics[scale=\scale,page=4]{addimages/\cGammaFile}
\includegraphics[scale=\scale,page=5]{addimages/\cGammaFile}
\includegraphics[scale=\scale,page=6]{addimages/\cGammaFile}
\hspace*{-0.2cm}\includegraphics[scale=\scale,page=7]{addimages/\cGammaFile}
\includegraphics[scale=0.86,page=\pagelegend]{addimages/\cGammaFile}
\caption{Leading powers $\hat{\Gamma}_i$ in $\gbar^2(1/a)$ for the $i$-th operator $\base_i^{(2)}$ of various $\ord(a)$-improved lattice discretisations plotted against their associated leading-order matching coefficients $\hat{c}_i$.
At subleading order, the value of the matching coefficient is unknown and instead a dashed line is drawn for the first subleading power, i.e.~at $\hat{\Gamma}_\text{min}+1$.
The faded line is drawn to highlight severely suppressed leading-order matching coefficients.
The gray lines indicate new massless contributions starting at NLO due to matching coefficients vanishing to leading order.
Here, such contributions may be more dominant than the ones at $\hat{\Gamma}_\text{min}+1$.
A case of particular interest are DWF with the two different choices of the domain wall height $M_5(g_0^2)=M_5^\text{TL}+\ord(g_0^2)$.}\label{fig:improvedActions}
\end{figure}
A more complicated situation is found for the $\ord(a)$-improved lattice actions.
Some common choices for 3-flavour QCD are depicted in figure~\ref{fig:improvedActions}.
Notice that both examples with DWF use the Iwasaki gauge action~\cite{Iwasaki:2011np}, while the examples for Overlap and $\ord(a)$-improved Wilson quarks use the L\"uscher-Weisz (LW) gauge action~\cite{Luscher:1984xn}.
Because there are many more operators, the spectrum $\hat{\Gamma}_i$ is much denser.
This will make it more difficult to decide which contribution dominates and may lead to complicated cancellations or pile-ups of different contributions.
Since DWF are (approximate) GW quarks, we find an identical spectrum $\hat{\Gamma}_i$ as for Overlap fermions here representing exact GW quarks.
Only $\ord(a)$-improved Wilson quarks have an additional set of chiral-symmetry breaking 4-fermion operators, but those contribute only at subleading powers to the spectrum due to vanishing TL matching coefficients.
Thus, in all cases chosen, the five leading powers are identical.
The massive $\hat{\Gamma}_\text{min}\approx -0.2\gg -3$, while in the massless case one finds $\hat{\Gamma}_\text{min}\approx 0.2\gg -3$.
Either way, the leading powers are fairly close to zero, which would correspond to the classical pure $a^2$ case and, in that respect, the theory is much better behaving than for the O(3) model.
The leading-order matching coefficients are not the same order of magnitude in the examples considered here and their hierarchy should be taken into account as well.

The most prominent example are (M\"obius) DWF with two different choices for the domain-wall height, see e.g.~\cite{Blum:2014tka,Nakayama:2016atf},
\begin{equation}
M_5(g_0^2)=M_5^\text{TL}+\ord(g_0^2)\,.
\end{equation}
Notice that $M_5^\text{TL}=1$ should be the ideal choice, but even in this case there are two coefficients severely enlarged by roughly a factor $\gtrsim 10$ compared to the coefficients, which should be dominant by simple power counting in $\gbar^2(1/a)$.
For $M_5^\text{TL}=1.8$, the leading three massive contributions become much larger and will likely be dominant unless the quark-mass suppression due to (very) small quark masses comes into play.
This particular choice shows that it is a good idea to tune the domain-wall height as a function of $g_0$ to arrive at $M_5^\text{TL}\simeq 1$ in the continuum limit $g_0\rightarrow 0$.

Overlap fermions have enlarged massive contributions with the largest matching coefficients close to DWF with $M_5^\text{TL}=1$.
Meanwhile, the massless contributions are less pronounced.
Especially the matching coefficient corresponding to $\hat{\Gamma}=1$ is much smaller.
Using the Wilson Dirac operator as the kernel for constructing Overlap fermions in combination with the LW gauge action, means that the leading-order matching coefficients of the massless contributions are identical to the case considered for $\ord(a)$-improved Wilson quarks.
The difference seen in the case of DWF stems purely from the aforementioned use of the Iwasaki gauge action.
Notably, for Wilson quarks the leading-order coefficients of the massive contributions are of the same magnitude as the massless ones.
Also the first two leading-order matching coefficients are suppressed by roughly a factor 10 compared to the largest leading-order matching coefficients, but the overall magnitude is smaller compared to, e.g., DWF.

As pointed out before, all lattice-fermion actions discussed here employ the Wilson Dirac operator in some way.
For those cases, tree-level $\ord(a^2)$ improvement can be achieved for all (massless) cases by choosing the LW action and replacing the Wilson lattice Dirac operator\footnote{Notice, that for $\ord(a)$-improved Wilson quarks, the Wilson Dirac operator involves the Sheikholeslami-Wohlert term~\cite{Sheikholeslami:1985ij} in contrast to the other fermion actions, where this term is optional.}, as suggested before~\cite{DeGrand:1995ji},
\begin{equation}
\hat{D}^\text{Wilson}\rightarrow\hat{D}^\text{Wilson}-\frac{a^2}{12}\sum_\mu \gamma_\mu\{\nabla_\mu+\nabla_\mu^*\}\nabla_\mu^*\nabla_\mu,
\end{equation}
where $\nabla_\mu$ and $\nabla_\mu^*$ are the covariant forward and backward lattice derivatives respectively.
With this replacement, all massless (red) lines in figure~\ref{fig:improvedActions} would be shifted up to and beyond the red-dashed line.
To achieve full Symanzik TL improvement at $\ord(a^2)$, also the massive contributions (including the SW term multiplied by quark masses in the Wilson case), would need to be tuned accordingly.
In the case of $\ord(a)$-improved Wilson quarks this additional improvement procedure would not necessarily gain us a full power in $\gbar^2(1/a)$ for the massless basis owing to the presence of chiral-symmetry breaking 4-fermion operators.
Those operators were suppressed but now might become the leading-power massless contributions.

\subsection{Effect of explicit \boldmath$\ord(a)$ improvement: $\Nf=2$ massless Wilson quarks in a finite volume}
While symmetries realised on the lattice ensure absence of operators in the SymEFT violating those symmetries, the same is not true for symmetries present only in the continuum theory.
Yet, continuum symmetry arguments can be used to show, e.g., absence of $\ord(a)$ lattice artifacts in maximally twisted mass QCD~\cite{Aoki:2006gh,Sint:2007ug} and similarly in massless Wilson QCD in a finite volume.

Focussing on the example of 2-flavour massless Wilson QCD in a finite volume brings us back to the minimal operator basis in eq.~\eqref{eq:dim5ops}, i.e.~the operator $\opFive_1$.
This operator is odd under the discrete symmetry transformation
\begin{equation}
\bar{\Psi}\rightarrow i\bar{\Psi}\gamma_5\tau^j,\quad \Psi\rightarrow i\gamma_5\tau^j\Psi,
\end{equation}
where $\tau^j$ is a Pauli matrix in flavour space.
The massless continuum QCD action is invariant.
As a consequence we find for any quantity $Q$ which is even under this transformation
\begin{equation}
\int\rmd^4z\,\langle Q\,\opFive_1(z)\rangle\rightarrow - \int\rmd^4z\,\langle Q\,\opFive_1(z)\rangle\,.
\end{equation}
Therefore, any $\ord(a)$ contribution from the operator $\opFive_1$ vanishes in finite volume for any quantity $Q$ with a non-zero continuum value.
This does not mean that the corresponding matching coefficient of $\opFive_1$ vanishes but the ``matrix element'' does.
Expanding the SymEFT to $\ord(a^2)$  will then yield the usual $\opSix_i$ operator insertions, but also a double insertion of $\opFive_1$ that transforms as
\begin{equation}
\int\rmd^4y\,\rmd^4z\,\langle Q\,\opFive_1(y)\opFive_1(z)\rangle\rightarrow\int\rmd^4y\,\rmd^4z\,\langle Q\,\opFive_1(y)\opFive_1(z)\rangle
\end{equation}
and therefore does not vanish.
Additionally, such double operator insertions give rise to contact terms, whose renormalisation affects the leading-order matching coefficients of the $\opSix_i$ operator basis~\cite{Husung:2022kvi}.
\begin{figure}
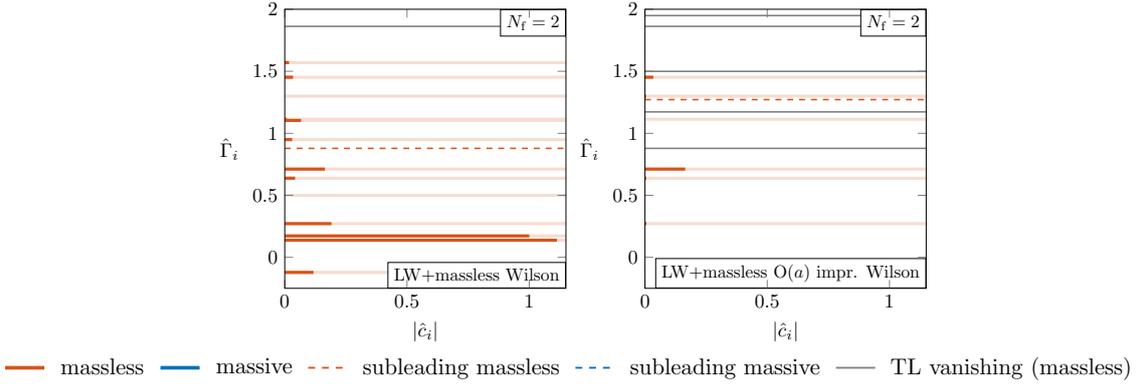
\centering
\includegraphics[scale=\scale,page=2]{addimages/\cGammaFile}
\includegraphics[scale=\scale,page=3]{addimages/\cGammaFile}\\
\includegraphics[scale=0.86,page=\pagelegend]{addimages/\cGammaFile}
\caption{Leading powers $\gbar^{2\hat{\Gamma}_i}(1/a)$ relevant for $\ord(a^2)$ corrections plotted against the leading-order matching coefficients for massless 2-flavour Wilson QCD with explicit $\ord(a)$ improvement (right) and without improvement (left).}\label{fig:improvement}
\end{figure}
The impact can be seen in figure~\ref{fig:improvement}.
For the unimproved case the spectrum of leading powers found is denser due to the presence of chiral-symmetry-breaking 4-fermion operators already at tree-level, which now have non-vanishing tree-level matching coefficients owing to the contact-term renormalisation.
The magnitude of two of those additional leading-order matching coefficients is also strongly enlarged.
On top of that, one of the chiral-symmetry breaking 4-fermion operators gives rise to $\hat{\Gamma}_\text{min}\approx -0.12$ compared to $\hat{\Gamma}_\text{min}\approx0.27$ in the explicitly $\ord(a)$-improved theory, where the matching coefficient associated to the latter is very suppressed $|\hat{c}_\text{min}|\approx 0.006$.

\section{Conclusion}
In all cases considered and moreover any choice of lattice QCD with $\Nf\leq 4$ flavours compatible with the symmetry constraints imposed here, the leading power $\gbar^{2\hat\Gamma_\mathrm{min}}(1/a)$ satisfies $\hat\Gamma_\mathrm{min}\gg -3$.
Coming back to the oversimplified example in figure~\ref{fig:exampleFakeData}, this is certainly good news and much better behaved than the O(3) model~\cite{Balog:2009np,Balog:2009yj}.
However, at $\ord(a^2)$ the large number of operators gives various contributions with powers $\hat{\Gamma}_i$ lying close to each other.
Here it is not clear which operator might contribute dominantly and one may find complicated lattice artifacts from cancellations and pile-ups.
On top of that, the corresponding leading-order matching coefficients $\hat{c}_i$ can have vastly different orders of magnitude.
This additional hierarchy must be taken into account, because it can undo relative suppressions in powers of $\gbar^2(1/a)$ making contributions dominant that otherwise would be assumed to be suppressed by simple power counting in $\gbar^2(1/a)$.

Having a look at the leading-order matching coefficients can also hint at whether (tree-level) Symanzik improvement is advisable.
We focussed here on three examples, namely the domain-wall height at tree-level $M_5(0)$ for DWF, tree-level $\ord(a^2)$ improvement of the lattice Wilson Dirac operator, as well as the use of automatic versus explicit $\ord(a)$ improvement of massless 2-flavour Wilson QCD in a finite volume.
But the general concept and arguments remain the same for other parameter choices or when relying on other continuum symmetries for automatic improvement.

While we cannot say with absolute certainty which operator is dominant, we still get an idea which contributions will likely be the most prominent ones due to low powers in $\gbar^2(1/a)$ and/or enlarged matching coefficients.
This knowledge should be taken into account when choosing ans\"atze for the continuum extrapolation of the form $a^2[2b_0\gbar^2(1/a)]^{\hat{\Gamma}}$ or more complicated choices.
Moreover, one should not take for granted that higher order corrections in the lattice spacing itself, i.e.~$\ord(a^{\nmin+1})$ or similar, are irrelevant nor that sufficiently small couplings are reached for a perturbative description, as presented here, to work well.
Lattice simulations nowadays start to reach couplings as small as $\gbar^2(1/a)/(4\pi)\sim0.21$.
The take-away message remains: be careful when doing continuum extrapolations.

Apart from the cases discussed here, a variety of cases is accessible now, including mixed actions or partially quenched actions~\cite{Husung:2022kvi}.
To allow further generalisation, the setup for the symbolic calculus can be accessed.\footnote{\url{https://github.com/nikolai-husung/Symanzik-QCD-workflow}}
An in-depth discussion of some common choices for $\Nf=3$ and $\Nf=4$ is presented in~\cite{Husung:2021mfl}.
The case of pure gauge has been considered before~\cite{Husung:2019ytz}, as well as the generalisation to Gradient flow, which requires an additional (third) operator in the minimal basis~\cite{Husung:2021tml} with $\hat{\gamma}=0$.

So far only spectral quantities have been considered, as they rely only on lattice artifacts arising from the lattice-discretised action.
For matrix elements of local fields, each local field involved will introduce its own minimal basis contributing even more powers in $\gbar^2(1/a)$ and associated matching coefficients.
A computation for local fermion bilinears of mass-dimension~3 is underway~\cite{H:inprepLocal}.

\textbf{Acknowledgements:} I thank Rainer Sommer and Peter Marquard for useful discussions and in particular for supervising my PhD project, which served as a starting point of this more general analysis.
In preparation of this plenary talk, I also got helpful suggestions from Rainer Sommer, Karl Jansen and Leonardo Chimirri.
Furthermore, I thank Hubert Simma, Kay Sch\"onwald and Agostino Patella for discussions and suggestions regarding the project,
as well as Andreas J\"uttner, Jonathan Flynn, and Rainer Sommer for checking the manuscript.

\bibliographystyle{JHEP}
\bibliography{lat22-husung.bbl}

\providecommand{\href}[2]{#2}\begingroup\raggedright\begin{thebibliography}{10}

\bibitem{Boyle:2017jwu}
P.~A. Boyle, L.~Del~Debbio, A.~J\"uttner, A.~Khamseh, F.~Sanfilippo and J.~T.
  Tsang, \emph{{The decay constants ${\mathbf{f_D}}$ and ${\mathbf{f_{D_{s}}}}$
  in the continuum limit of ${\mathbf{N_f=2+1}}$ domain wall lattice QCD}},
  \href{https://doi.org/10.1007/JHEP12(2017)008}{\emph{JHEP} {\bfseries 12}
  (2017) 008} [\href{https://arxiv.org/abs/1701.02644}{{\ttfamily
  1701.02644}}].

\bibitem{Bruno:2017gxd}
{\scshape ALPHA} collaboration, \emph{{QCD Coupling from a Nonperturbative
  Determination of the Three-Flavor $\Lambda$ Parameter}},
  \href{https://doi.org/10.1103/PhysRevLett.119.102001}{\emph{Phys. Rev. Lett.}
  {\bfseries 119} (2017) 102001}
  [\href{https://arxiv.org/abs/1706.03821}{{\ttfamily 1706.03821}}].

\bibitem{Husung:2019ytz}
N.~Husung, P.~Marquard and R.~Sommer, \emph{{Asymptotic behavior of cutoff
  effects in Yang-Mills theory and in Wilson's lattice QCD}},
  \href{https://doi.org/10.1140/epjc/s10052-020-7685-4}{\emph{Eur.\ Phys.\ J.\
  C} {\bfseries 80} (2020) 200}
  [\href{https://arxiv.org/abs/1912.08498}{{\ttfamily 1912.08498}}].

\bibitem{Husung:2021tml}
N.~Husung, \emph{{Logarithmic corrections in Symanzik\textquoteright{}s
  effective theory of lattice QCD}}, Ph.D. thesis, Humboldt U., Berlin, 8,
  2021.
\newblock 10.18452/22944.

\bibitem{Husung:2021mfl}
N.~Husung, P.~Marquard and R.~Sommer, \emph{{The asymptotic approach to the
  continuum of lattice QCD spectral observables}},
  \href{https://doi.org/10.1016/j.physletb.2022.137069}{\emph{Phys. Lett. B}
  {\bfseries 829} (2022) 137069}
  [\href{https://arxiv.org/abs/2111.02347}{{\ttfamily 2111.02347}}].

\bibitem{Wilson:1974}
K.~G. Wilson, \emph{Confinement of quarks},
  \href{https://doi.org/10.1103/PhysRevD.10.2445}{\emph{Phys. Rev. D}
  {\bfseries 10} (1974) 2445}.

\bibitem{Wilson:1975id}
K.~G. Wilson, \emph{{Quarks and Strings on a Lattice}},  in \emph{{New
  Phenomena in Subnuclear Physics: Proceedings, International School of
  Subnuclear Physics, Erice, Sicily, Jul 11-Aug 1 1975. Part A}}, p.~99, 1975.

\bibitem{Luscher:1996ug}
M.~{L\"uscher}, S.~Sint, R.~Sommer, P.~Weisz and U.~Wolff,
  \emph{{Nonperturbative O(a) improvement of lattice QCD}},
  \href{https://doi.org/10.1016/S0550-3213(97)00080-1}{\emph{Nucl. Phys.}
  {\bfseries B491} (1997) 323}
  [\href{https://arxiv.org/abs/hep-lat/9609035}{{\ttfamily hep-lat/9609035}}].

\bibitem{Ginsparg:1981bj}
P.~H. Ginsparg and K.~G. Wilson, \emph{{A Remnant of Chiral Symmetry on the
  Lattice}}, \href{https://doi.org/10.1103/PhysRevD.25.2649}{\emph{Phys. Rev.}
  {\bfseries D25} (1982) 2649}.

\bibitem{Neuberger:1997fp}
H.~Neuberger, \emph{{Exactly massless quarks on the lattice}},
  \href{https://doi.org/10.1016/S0370-2693(97)01368-3}{\emph{Phys.\ Lett.\ B}
  {\bfseries 417} (1998) 141}
  [\href{https://arxiv.org/abs/hep-lat/9707022}{{\ttfamily hep-lat/9707022}}].

\bibitem{Neuberger:1998wv}
H.~Neuberger, \emph{{More about exactly massless quarks on the lattice}},
  \href{https://doi.org/10.1016/S0370-2693(98)00355-4}{\emph{Phys.\ Lett.\ B}
  {\bfseries 427} (1998) 353}
  [\href{https://arxiv.org/abs/hep-lat/9801031}{{\ttfamily hep-lat/9801031}}].

\bibitem{Kaplan:1992bt}
D.~B. Kaplan, \emph{{A Method for simulating chiral fermions on the lattice}},
  \href{https://doi.org/10.1016/0370-2693(92)91112-M}{\emph{Phys. Lett. B}
  {\bfseries 288} (1992) 342}
  [\href{https://arxiv.org/abs/hep-lat/9206013}{{\ttfamily hep-lat/9206013}}].

\bibitem{Furman_1995}
V.~Furman and Y.~Shamir, \emph{Axial symmetries in lattice qcd with kaplan
  fermions}, \href{https://doi.org/10.1016/0550-3213(95)00031-m}{\emph{Nuclear
  Physics B} {\bfseries 439} (1995) 54–78}.

\bibitem{Brower:2012vk}
R.~C. Brower, H.~Neff and K.~Orginos, \emph{{The M\"obius domain wall fermion
  algorithm}}, \href{https://doi.org/10.1016/j.cpc.2017.01.024}{\emph{Comput.
  Phys. Commun.} {\bfseries 220} (2017) 1}
  [\href{https://arxiv.org/abs/1206.5214}{{\ttfamily 1206.5214}}].

\bibitem{Luscher:1998pqa}
M.~{L\"uscher}, \emph{{Exact chiral symmetry on the lattice and the
  Ginsparg-Wilson relation}},
  \href{https://doi.org/10.1016/S0370-2693(98)00423-7}{\emph{Phys. Lett. B}
  {\bfseries 428} (1998) 342}
  [\href{https://arxiv.org/abs/hep-lat/9802011}{{\ttfamily hep-lat/9802011}}].

\bibitem{Symanzik:1979ph}
K.~Symanzik, \emph{{Cutoff dependence in lattice $\phi_4^4$ theory}},
  \href{https://doi.org/10.1007/978-1-4684-7571-5_18}{\emph{NATO Sci. Ser. B}
  {\bfseries 59} (1980) 313}.

\bibitem{Symanzik:1981hc}
K.~Symanzik, \emph{{Some Topics in Quantum Field Theory}},  in
  \emph{{Mathematical Problems in Theoretical Physics. Proceedings, 6th
  International Conference on Mathematical Physics, West Berlin, Germany,
  August 11-20, 1981}}, pp.~47--58, 1981.

\bibitem{Symanzik:1983dc}
K.~Symanzik, \emph{{Continuum Limit and Improved Action in Lattice Theories. 1.
  Principles and $\phi^4$ Theory}},
  \href{https://doi.org/10.1016/0550-3213(83)90468-6}{\emph{Nucl. Phys.}
  {\bfseries B226} (1983) 187}.

\bibitem{Symanzik:1983gh}
K.~Symanzik, \emph{{Continuum Limit and Improved Action in Lattice Theories. 2.
  O(N) Nonlinear Sigma Model in Perturbation Theory}},
  \href{https://doi.org/10.1016/0550-3213(83)90469-8}{\emph{Nucl. Phys.}
  {\bfseries B226} (1983) 205}.

\bibitem{Balog:2009np}
J.~Balog, F.~Niedermayer and P.~Weisz, \emph{{The Puzzle of apparent linear
  lattice artifacts in the 2d non-linear sigma-model and Symanzik's solution}},
  \href{https://doi.org/10.1016/j.nuclphysb.2009.09.007}{\emph{Nucl. Phys.}
  {\bfseries B824} (2010) 563}
  [\href{https://arxiv.org/abs/0905.1730}{{\ttfamily 0905.1730}}].

\bibitem{Balog:2009yj}
J.~Balog, F.~Niedermayer and P.~Weisz, \emph{{Logarithmic corrections to
  O($a^2$) lattice artifacts}},
  \href{https://doi.org/10.1016/j.physletb.2009.04.082}{\emph{Phys. Lett.}
  {\bfseries B676} (2009) 188}
  [\href{https://arxiv.org/abs/0901.4033}{{\ttfamily 0901.4033}}].

\bibitem{Luthe:2017ttc}
T.~Luthe, A.~Maier, P.~Marquard and Y.~{Schr\"oder}, \emph{{Complete
  renormalization of QCD at five loops}},
  \href{https://doi.org/10.1007/JHEP03(2017)020}{\emph{JHEP} {\bfseries 03}
  (2017) 020} [\href{https://arxiv.org/abs/1701.07068}{{\ttfamily
  1701.07068}}].

\bibitem{Luscher:1996sc}
M.~{L\"uscher}, S.~Sint, R.~Sommer and P.~Weisz, \emph{{Chiral symmetry and
  O(a) improvement in lattice QCD}},
  \href{https://doi.org/10.1016/0550-3213(96)00378-1}{\emph{Nucl. Phys.}
  {\bfseries B478} (1996) 365}
  [\href{https://arxiv.org/abs/hep-lat/9605038}{{\ttfamily hep-lat/9605038}}].

\bibitem{Sheikholeslami:1985ij}
B.~Sheikholeslami and R.~Wohlert, \emph{{Improved Continuum Limit Lattice
  Action for QCD with Wilson Fermions}},
  \href{https://doi.org/10.1016/0550-3213(85)90002-1}{\emph{Nucl. Phys.}
  {\bfseries B259} (1985) 572}.

\bibitem{Husung:2022kvi}
N.~Husung, \emph{{Logarithmic corrections to O($a$) and O($a^2$) effects in
  lattice QCD with Wilson or Ginsparg-Wilson quarks}},
  \href{https://arxiv.org/abs/2206.03536}{{\ttfamily 2206.03536}}.

\bibitem{tHooft:1975uxh}
G.~'t~Hooft, \emph{{The Background Field Method in Gauge Field Theories}},  in
  \emph{{Functional and Probabilistic Methods in Quantum Field Theory. 1.
  Proceedings, 12th Winter School of Theoretical Physics, Karpacz, Feb 17-March
  2, 1975}}, pp.~345--369, 1975.

\bibitem{Abbott:1980hw}
L.~F. Abbott, \emph{{The Background Field Method Beyond One Loop}},
  \href{https://doi.org/10.1016/0550-3213(81)90371-0}{\emph{Nucl. Phys.}
  {\bfseries B185} (1981) 189}.

\bibitem{Abbott:1981ke}
L.~F. Abbott, \emph{{Introduction to the Background Field Method}}, {\emph{Acta
  Phys. Polon.} {\bfseries B13} (1982) 33}.

\bibitem{Luscher:1995vs}
M.~{L\"uscher} and P.~Weisz, \emph{{Background field technique and
  renormalization in lattice gauge theory}},
  \href{https://doi.org/10.1016/0550-3213(95)00346-T}{\emph{Nucl. Phys.}
  {\bfseries B452} (1995) 213}
  [\href{https://arxiv.org/abs/hep-lat/9504006}{{\ttfamily hep-lat/9504006}}].

\bibitem{tHooft:1972tcz}
G.~'t~Hooft and M.~J.~G. Veltman, \emph{{Regularization and Renormalization of
  Gauge Fields}},
  \href{https://doi.org/10.1016/0550-3213(72)90279-9}{\emph{Nucl. Phys.}
  {\bfseries B44} (1972) 189}.

\bibitem{tHooft:1973mfk}
G.~'t~Hooft, \emph{{Dimensional regularization and the renormalization group}},
  \href{https://doi.org/10.1016/0550-3213(73)90376-3}{\emph{Nucl. Phys.}
  {\bfseries B61} (1973) 455}.

\bibitem{Bardeen:1978yd}
W.~A. Bardeen, A.~J. Buras, D.~W. Duke and T.~Muta, \emph{{Deep Inelastic
  Scattering Beyond the Leading Order in Asymptotically Free Gauge Theories}},
  \href{https://doi.org/10.1103/PhysRevD.18.3998}{\emph{Phys. Rev.} {\bfseries
  D18} (1978) 3998}.

\bibitem{Misiak:1994zw}
M.~Misiak and M.~{M\"unz}, \emph{{Two loop mixing of dimension five flavor
  changing operators}},
  \href{https://doi.org/10.1016/0370-2693(94)01553-O}{\emph{Phys. Lett.}
  {\bfseries B344} (1995) 308}
  [\href{https://arxiv.org/abs/hep-ph/9409454}{{\ttfamily hep-ph/9409454}}].

\bibitem{Chetyrkin:1997fm}
K.~G. Chetyrkin, M.~Misiak and M.~{M\"unz}, \emph{{Beta functions and anomalous
  dimensions up to three loops}},
  \href{https://doi.org/10.1016/S0550-3213(98)00122-9}{\emph{Nucl. Phys.}
  {\bfseries B518} (1998) 473}
  [\href{https://arxiv.org/abs/hep-ph/9711266}{{\ttfamily hep-ph/9711266}}].

\bibitem{Iwasaki:2011np}
Y.~Iwasaki, \emph{{Renormalization Group Analysis of Lattice Theories and
  Improved Lattice Action. II. Four-dimensional non-Abelian SU(N) gauge
  model}},  \href{https://arxiv.org/abs/1111.7054}{{\ttfamily 1111.7054}}.

\bibitem{Luscher:1984xn}
M.~{L\"uscher} and P.~Weisz, \emph{{On-Shell Improved Lattice Gauge Theories}},
  \href{https://doi.org/10.1007/BF01206178}{\emph{Commun.\ Math.\ Phys.}
  {\bfseries 97} (1985) 59}.

\bibitem{Blum:2014tka}
{\scshape RBC, UKQCD} collaboration, \emph{{Domain wall QCD with physical quark
  masses}}, \href{https://doi.org/10.1103/PhysRevD.93.074505}{\emph{Phys. Rev.
  D} {\bfseries 93} (2016) 074505}
  [\href{https://arxiv.org/abs/1411.7017}{{\ttfamily 1411.7017}}].

\bibitem{Nakayama:2016atf}
K.~Nakayama, B.~Fahy and S.~Hashimoto, \emph{{Short-distance charmonium
  correlator on the lattice with M\"obius domain-wall fermion and a
  determination of charm quark mass}},
  \href{https://doi.org/10.1103/PhysRevD.94.054507}{\emph{Phys. Rev. D}
  {\bfseries 94} (2016) 054507}
  [\href{https://arxiv.org/abs/1606.01002}{{\ttfamily 1606.01002}}].

\bibitem{DeGrand:1995ji}
T.~A. DeGrand, A.~Hasenfratz, P.~Hasenfratz and F.~Niedermayer, \emph{{The
  Classically perfect fixed point action for SU(3) gauge theory}},
  \href{https://doi.org/10.1016/0550-3213(95)00458-5}{\emph{Nucl. Phys. B}
  {\bfseries 454} (1995) 587}
  [\href{https://arxiv.org/abs/hep-lat/9506030}{{\ttfamily hep-lat/9506030}}].

\bibitem{Aoki:2006gh}
S.~Aoki and O.~{B\"ar}, \emph{{Automatic O(a) improvement for twisted-mass
  QCD}}, \href{https://doi.org/10.22323/1.032.0165}{\emph{PoS} {\bfseries
  LAT2006} (2006) 165} [\href{https://arxiv.org/abs/hep-lat/0610098}{{\ttfamily
  hep-lat/0610098}}].

\bibitem{Sint:2007ug}
S.~Sint, \emph{{Lattice QCD with a chiral twist}},  in \emph{{Workshop on
  Perspectives in Lattice QCD Nara, Japan, October 31-November 11, 2005}},
  2007, \href{https://arxiv.org/abs/hep-lat/0702008}{{\ttfamily
  hep-lat/0702008}}, \href{https://doi.org/10.1142/9789812790927_0004}{DOI}.

\bibitem{H:inprepLocal}
N.~Husung, \emph{{O($a^2$) lattice artifacts for local fermion bilinears}},
  {\emph{{in preparation}} }.

\end{thebibliography}\endgroup

\end{document}